\documentclass[conference,10pt]{IEEEtran}

\usepackage[utf8]{inputenc}
\usepackage{graphicx}
\usepackage{cite}
\usepackage{amsmath,amssymb,amsfonts}
\usepackage{algorithmic}
\usepackage{textcomp}
\usepackage{xcolor}
\usepackage{microtype}
\usepackage{booktabs}
\usepackage{placeins} 
\usepackage[pdfencoding=auto, psdextra, breaklinks]{hyperref} 
\usepackage[capitalise]{cleveref}
\usepackage{optidef}
\usepackage{bm}
\usepackage{siunitx}
\usepackage{multicol,multirow}

\usepackage{tikz}
\usepackage{makecell} 

\def\BibTeX{{\rm B\kern-.05em{\sc i\kern-.025em b}\kern-.08em
    T\kern-.1667em\lower.7ex\hbox{E}\kern-.125emX}}

\hypersetup{
    colorlinks,
    linkcolor={red!90!black},
    citecolor={blue!90!black},
    urlcolor={blue!90!black}
}

\newcommand{\ie}{i.e., }

\renewcommand{\v}[1]{\bm{#1}}  
\def\cycSize{C_k}  
\def\ModSet{\mathcal{M}} 
\def\modf{\mu} 
\def\cidx{c}  

\def\csizeest{Q}  

\DeclareMathSymbol{\shortminus}{\mathbin}{AMSa}{"39}
\newcommand{\medminus}{\scalebox{0.6}[0.7]{\(-\)}}
\newcommand{\minus}{\mathchoice{-}{-}{\medminus}{\shortminus}}
\renewcommand{\baselinestretch}{0.98}
\usepackage[font=small,position=bottom,belowskip=0.1pt,aboveskip=0.1pt]{caption}
\usepackage[font=small,position=bottom,belowskip=0.1pt,aboveskip=0.1pt]{subcaption}


\begin{document}

\title{A two-step approach for speech enhancement in low-SNR scenarios using cyclostationary beamforming and DNNs}
\author{
%
\IEEEauthorblockN{
Giovanni Bologni\textsuperscript{1*}, 
Nicol\'as Arrieta Larraza\textsuperscript{2*}, 
Richard Heusdens\textsuperscript{1} 
and Richard C.~Hendriks\textsuperscript{1}}
\IEEEauthorblockA{\textit{
\textsuperscript{1}Delft University of Technology, EEMCS faculty}, Delft, the Netherlands, \{G.Bologni, R.Heusdens, R.C.Hendriks\}@tudelft.nl}
\IEEEauthorblockA{\textit{
\textsuperscript{2}Bang \& Olufsen}, Lyngby, Denmark, nial@bang-olufsen.dk}
}
\maketitle
\renewcommand{\thefootnote}{\fnsymbol{footnote}}
\footnotetext[1]{These authors contributed equally.}
\begin{abstract}
Deep Neural Networks (DNNs) often struggle to suppress noise at low signal-to-noise ratios (SNRs). 
This paper addresses speech enhancement in scenarios dominated by harmonic noise and proposes a framework that integrates cyclostationarity-aware preprocessing with lightweight DNN-based denoising.
A cyclic minimum power distortionless response (cMPDR) spectral beamformer is used as a preprocessing block. It exploits the spectral correlations of cyclostationary noise to suppress harmonic components prior to learning-based enhancement and does not require modifications to the DNN architecture.
The proposed pipeline is evaluated in a single-channel setting using two DNN architectures: a simple and lightweight convolutional recurrent neural network (CRNN), and a state-of-the-art model, namely ultra-low complexity network (ULCNet).
Experiments on synthetic data and real-world recordings dominated by rotating machinery noise demonstrate consistent improvements over end-to-end DNN baselines, particularly at low SNRs.
Remarkably, a parameter-efficient CRNN with cMPDR preprocessing surpasses the performance of the larger ULCNet operating on raw or Wiener-filtered inputs.
These results indicate that explicitly incorporating cyclostationarity as a signal prior is more effective than increasing model capacity alone for suppressing harmonic interference.
\end{abstract}


\section{Introduction}\noindent
Reliable speech communication in high-noise industrial environments remains a challenge for hearables and electronic hearing protection devices \cite{zaki_method_2025}.
While Deep Neural Networks (DNNs) excel in general denoising tasks \cite{cohen_explainable_2025}, they become unreliable at low signal-to-noise ratios (SNRs), often introducing speech distortion or failing to suppress noise \cite{hao_unetgan_2019,schulz_effects_2026}.
To alleviate this, two-step pipelines preprocess the input prior to DNN-based enhancement \cite{hao_masking_2020,wang_lightweight_2025}, while other methods incorporate prior knowledge on the harmonic structure of speech \cite{le_harmonic_2023,hou_snr-progressive_2025}.
In contrast, the statistical properties of the noise are often neglected.

\begin{figure}[tbph]
    \centering
    \input{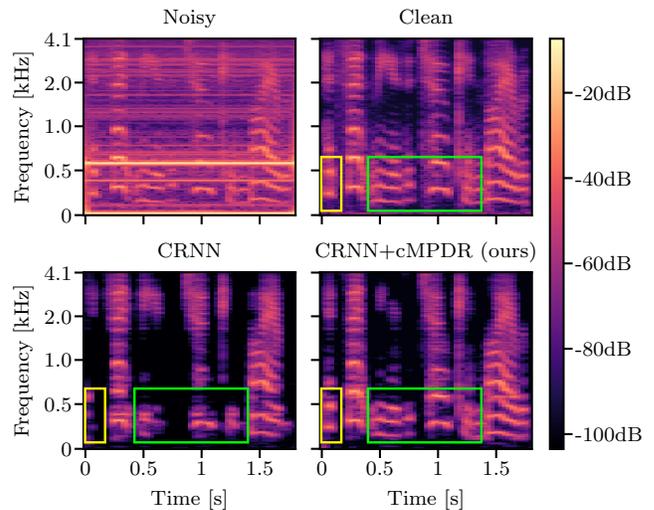}
    \caption{Mel spectrograms of a recording from the IDMT-ISA Electric Engine dataset, showing noisy input, clean target, and outputs of the CRNN and proposed CRNN+cMPDR pipeline. The highlights show that CRNN+cMPDR can recover heavily masked speech components, illustrating the benefits of cyclostationarity-aware preprocessing.}
    \label{fig:spectrograms}
\end{figure}

Noise generated by rotating machinery exhibits \textit{cyclostationarity}, combining random fluctuations with deterministic periodic components \cite{antoni_cyclostationarity_2009}.
This work focuses on such noise due to its strong regularity and practical relevance.
In the frequency domain, cyclostationarity manifests as spectral correlations between harmonically related frequency bins.
Standard short-time Fourier transform (STFT)-based DNNs are ill-suited to exploit this structure.
First, most architectures operate on magnitude spectra, discarding the inter-frequency phase relationships that characterize cyclostationarity \cite{gardner_measurement_1986}.
Second, even complex-valued networks are affected by the frequency- and frame-dependent time-shift introduced by the STFT, which destroys coherent phase relationships across frequency bins.
While it is possible to partially compensate for these shifts \cite[Eq.~10]{bologni_wideband_2025}, exact correction requires explicit knowledge of the underlying signal frequencies \cite{krawczyk_stft_2014}.
As a result, standard DNNs do not reliably suppress harmonic noise, as they lack the inductive bias required to model these spectral couplings.
Consistent with this limitation, our experiments show that increasing DNN capacity alone does not enable reliable exploitation of cyclostationary structure, even when trained on tens of hours of data with highly regular harmonic interference.

In this paper, we propose a single-channel speech enhancement pipeline that integrates cyclostationarity-informed preprocessing to address this limitation.
We combine a cyclic minimum power distortionless response (cMPDR) beamformer \cite{bologni_mvdr_2025} with a masking-based DNN.
While the cMPDR is applicable to multichannel arrays, we focus on the single-channel formulation to demonstrate that exploiting spectral redundancy alone is sufficient to suppress interference, even without spatial diversity.
The cMPDR, originally introduced in a separate work \cite{bologni_mvdr_2025}, is used here as a front end that can be paired without modification with different DNN architectures.
It operates as a single-channel spectral beamformer, linearly combining frequency-shifted versions of the signal to suppress harmonic interferences while preserving the target speech \cite{bologni_cyclic_2025-2}.
Crucially, frequency shifting is performed through time-domain modulation prior to STFT analysis, ensuring precise alignment of cyclostationary components that is not limited by the STFT resolution.
The cMPDR output is then processed by the DNN, allowing the learning-based stage to build upon the gains provided by cyclostationarity-aware preprocessing rather than attempting to infer this structure implicitly.
This hybrid approach is evaluated using two lightweight DNN architectures, a convolutional recurrent neural network (CRNN) and the state-of-the-art ultra-low complexity network (ULCNet) \cite{shetu_ultra_2024}, on both synthetic and real-world data.
Results show consistent improvements over DNN-only baselines, particularly in low-SNR conditions with strong harmonic interference.
Notably, while the DNN can readily replace conventional preprocessing such as Wiener filtering, its performance gains are largely complementary to those provided by cMPDR, rather than redundant.

Our contributions are threefold:
$(i)$ we introduce a hybrid framework integrating cMPDR-based preprocessing with DNN-based speech enhancement; 
$(ii)$ we show that a small DNN with cMPDR preprocessing can outperform a larger model operating on raw or Wiener-filtered data, achieving superior interference suppression with approximately 30\% fewer parameters;
$(iii)$ we provide empirical evidence that increasing DNN capacity alone does not recover the gains achieved by cyclostationarity-aware preprocessing, confirming the necessity of specialized inductive biases for harmonic noise removal. 
Additionally, the code is publicly available on Github\footnote{\url{https://github.com/narrietal/cMPDR_DNN}}.


\section{Proposed method: cMPDR-DNN}\noindent
The noisy measurements in the STFT domain for a single microphone are given by:
\begin{align}\label{eq:sig_mod_nb}
x(\omega_k, \ell) &= d(\omega_k, \ell) + v(\omega_k, \ell),
\end{align}
where $k=0,\ldots,K-1$ and $\ell=0,\ldots,L-1$ denote the frequency and time indices, respectively, $\omega_k = 2\pi k / K$ is the angular frequency, $d(\omega_k, \ell)$ is the target speech, and $v(\omega_k, \ell)$ is the additive noise.
For brevity, indices are omitted where the context is unambiguous.
The proposed approach is a two-stage enhancement procedure designed for low-SNR scenarios with harmonic noise.
To exploit the spectral redundancy of the noise, we first define an augmented observation vector $\v{x}(\ModSet_k, \omega_k, \ell) \in \mathbb{C}^{\cycSize}$ formed by stacking frequency-shifted copies of the input signal:
\begin{multline}\label{eq:augmented_vector}
\v{x}(\ModSet_k, \omega_k, \ell) = \\
\begin{bmatrix*}
x(\omega_k, \ell) &
x(\omega_k \minus \modf_{1_k}, \ell) &
\cdots &
x(\omega_k \minus \modf_{\cycSize \minus 1}, \ell)
\end{bmatrix*}^T,
\end{multline}
where $(\cdot)^T$ denotes transposition and
$\ModSet_k = \{\modf_{\cidx_k}\}_{\cidx_k = 0}^{\cycSize - 1}$
is the \textit{modulation set} containing the applied frequency shifts. 
It is important to note that the shifts $\modf_{\cidx_k}$ correspond to the physical cyclic frequencies of the noise and generally do not align with the discrete STFT bin centers. 
Consequently, the shifted components $x(\omega_k - \modf_{\cidx_k})$ are generally not accessible via standard indexing of the STFT matrix. 
The augmented vector $\v{x}$ explicitly constructs these off-grid components via time-domain modulation, recovering fine-grained spectral correlations that are otherwise lost due to the fixed STFT resolution.

The intermediate signal $y$ is the output of the first stage of enhancement, a preprocessor $\Lambda \in \{\Lambda_{\text{cMPDR}}, \Lambda_{\text{Wiener}}, \Lambda_{\text{Id}}\}$ acting on the augmented vector:
\begin{align}
    y(\omega_k, \ell) = \Lambda(\v{x}(\ModSet_k, \omega_k, \ell)).
\end{align}
For the baseline Wiener filter and the identity mapping ($\Lambda_{\text{Wiener}}$ and $\Lambda_{\text{Id}}$, \cref{ssec:baselines-metrics}), the modulation set is trivial, \ie $\ModSet_k = \{0\}$, and $\v{x}$ reduces to the scalar noisy input $x(\omega_k)$. 
For the cMPDR ($\Lambda_{\text{cMPDR}}$, \cref{ssec:cmpdr}), $\ModSet_k$ also contains the non-zero cyclic frequencies estimated from the data, allowing the beamformer to exploit statistical correlations across the continuous frequency spectrum.

Finally, 
the preprocessed signal for all frequency bins and time frames is collected into a matrix $\v{Y} \in \mathbb{C}^{K \times L}$. 
A DNN $\Phi \in \{\Phi_{\text{CRNN}}, \Phi_{\text{ULCNet}}\}$ suppresses the residual noise and produces a speech estimate $\v{\hat{D}}$:
\begin{equation}
    \v{\hat{D}} = \v{M} \odot \v{Y},
\end{equation}
where $\odot$ denotes the Hadamard product and $\v{M} = \Phi(\v{Y})$ is the estimated mask.
For $\Phi_{\text{CRNN}}$, $\v{M} \in \mathbb{R}_{+}^{K \times L}$ is a real-valued magnitude mask $M_\text{r}(|\v{Y}|)$, while for $\Phi_{\text{ULCNet}}$, $\v{M} \in \mathbb{C}^{K \times L}$ is a complex-valued mask $M_\text{c}(\v{Y})$ performing joint magnitude and phase refinement.

\subsection{cMPDR preprocessor 
\texorpdfstring{($\Lambda_{\text{cMPDR}}$)}{}
}\label{ssec:cmpdr}\noindent
To perform the initial stage of enhancement, we employ the cyclostationarity-aware cMPDR beamformer \cite{bologni_mvdr_2025}.
By treating frequency-shifted versions of a single-channel signal as virtual channels, the cMPDR exploits the spectral redundancy characteristic of harmonic noise. 
This stage relies on the spectral covariance matrix $\v{S}_{\v{x}} \in \mathbb{C}^{\cycSize\times \cycSize}$, which  
is estimated by recursive temporal averaging of the noisy recording \cite{antoni_cyclostationarity_2009}:
\begin{multline}\label{eq:spectral_cov_definition}
    \hat{\v{S}}_{\v{x}}(\ModSet_k, \omega_k, \ell) \leftarrow \beta_x \hat{\v{S}}_{\v{x}}(\ModSet_k, \omega_k, \ell \minus 1) + \\
    (1 - \beta_x) \v{x}(\ModSet_k, \omega_k, \ell)\v{x}(\ModSet_k, \omega_k, \ell)^H,
\end{multline}
where $(\cdot)^H$ denotes conjugate transposition and $\beta_x=0.95$ is a smoothing constant.
The cMPDR filter $\v{w} \in \mathbb{C}^{\cycSize}$ is designed to minimize the empirical output power while preserving the non-modulated component:
\begin{mini}
    {\v{w}}{\v{w}^H \hat{\v{S}}_{\v{x}}(\ModSet_k, \omega_k, \ell) \v{w}}
    {}
    {}{}
    \addConstraint{\v{w}^H \v{e}_{1}}{= 1},
    \label{eq:cmvdr_problem}
\end{mini}
where $\v{e}_1 = [1, 0, \cdots, 0]^T$. 
The output for a single frequency-bin and time-frame is computed in closed-form as:
\begin{align}\label{eq:cmvdr_solution}
y = \Lambda_{\text{cMPDR}}(\v{x}) = \left(\frac{\hat{\v{S}}_{\v{x}}^{-1}\v{e}_1}{\v{e}_1^H \hat{\v{S}}_{\v{x}}^{-1} \v{e}_1}\right)^H \v{x}.
\end{align}
The effectiveness of this filtering depends on the modulation set $\ModSet_k$. 
To maximize correlations between the virtual channels, the frequency shifts $\modf_{\cidx_k}$ must align with the cyclic frequencies of the noise.
Including uncorrelated channels is theoretically useless and can be detrimental in the presence of estimation errors.
Following \cite{bologni_mvdr_2025}, we assume the noise resonant frequencies remain constant over the recording duration. 
Thus, $\ModSet_k$ is estimated once per recording by identifying the top $\csizeest$ peaks of the periodogram \cite{stoica_spectral_2005}.
Candidate modulations are determined as pairwise frequency differences between peak locations, and the final set $\ModSet_k$ is pruned to include only those candidates exhibiting high spectral coherence with the unshifted input signal, yielding a set of cardinality $\cycSize$.
While this batch estimation is suitable for stable mechanical noise, adaptive tracking of time-varying resonances is reserved for future work.
By suppressing dominant harmonic interference in this first stage, the cMPDR effectively reduces the mapping burden for the subsequent DNN, enabling robust performance even with very low-complexity architectures.

\subsection{Deep learning architectures}\noindent
In the proposed pipeline, the preprocessed signal is further processed by a DNN to produce an enhanced target speech estimate.
We consider two low-complexity DNN architectures: a simple CRNN and a state-of-the-art model, ULCNet.
\subsubsection{CRNN 
\texorpdfstring{($\Phi_{\text{CRNN}}$)}{}}
\label{ssec:crnn}\noindent
This architecture operates in the STFT domain. 
It takes the magnitude spectrum as input and it outputs a real-valued time-frequency mask. 
The architecture consists of three 2D convolutional layers for feature extraction, followed by a Gated Recurrent Unit (GRU) layer to capture long-range temporal dependencies across time frames. 
A final fully connected layer generates the mask $M_\text{r}(|\v{Y}|)$.
\subsubsection{ULCNet (\texorpdfstring{$\Phi_{\text{ULCNet}}$}{})}
\label{ssec:ulc}\noindent
This model also operates in the time-frequency domain and comprises two stages.
In the first stage, a modified power-law compression is applied to both the real and imaginary components of the input STFT.
Subsequently, a channel-wise feature reorientation mechanism is employed to reduce the dimensionality of the input features, thereby improving computational efficiency. 
The processed features are then passed through a sequence of depthwise separable convolutional layers that function as feature extractors. 
To expand the receptive field, a bidirectional GRU layer operating along the frequency axis is applied. 
Additionally, subband-level temporal GRU-based units are integrated to enhance spectral modeling capabilities. This stage concludes with the estimation of a real-valued magnitude mask through two fully connected layers.
The second stage focuses on phase refinement. 
Lastly, a convolutional neural network (CNN) processes intermediate representations derived from the estimated magnitude mask and the noisy phase to obtain a final complex mask $M_\text{c}(\v{Y})$.

%
%

\section{Experimental setup}\noindent
This section describes the datasets, baseline models, performance metrics, and implementation details used in our evaluation.

\subsection{Datasets}\noindent
\subsubsection{Synthetic dataset}\label{ssec:synthetic_data}\noindent
We generate harmonic noise signals and mix them with clean speech from the 2020 Deep Noise Suppression (DNS) Challenge dataset \cite{reddy_interspeech_2020-1} at a sampling rate of $\SI{16}{\kilo\hertz}$ with SNR values randomly drawn from a uniform distribution ranging from $\SI{-20}{\decibel}$ to $\SI{0}{\decibel}$ to obtain a 50-hour dataset.
%
Noise follows the random harmonic CS model from \cite{bologni_mvdr_2025}, where spectral correlation is set to $\beta=0.9$, the number of harmonic components is set to $P_u=10$, and the fundamental frequency is drawn randomly from $\mathcal{U}(\SI{60}{\hertz},\SI{150}{\hertz})$.


\begin{table*}[tb]
\centering
\caption{Results using noise from the DNS Challenge 2020 (DNS) and IDMT-ISA Electric Engine dataset (IDMT). Mean SI-SDR [\si{\decibel}], DNSMOS (DMOS), and STOI are reported for SNR ranges $[-20,-10)$ \si{\decibel} and $[-10,0]$ \si{\decibel}. Model parameters are reported in millions.}
\label{tab:enhancement_comparison}
\setlength{\tabcolsep}{5.8pt}  
\footnotesize
\begin{tabular}{ccc ccc ccc ccc ccc}
\toprule
\multirow{3}{*}{Model} &
\multirow{3}{*}{\makecell{Params}} &
\multirow{3}{*}{Preproc.} &
\multicolumn{6}{c}{IDMT} &
\multicolumn{6}{c}{DNS} \\
\cmidrule(lr){4-9} \cmidrule(lr){10-15}
& & &
\multicolumn{3}{c}{SNR $[-20,-10)$} &
\multicolumn{3}{c}{SNR $[-10,0]$} &
\multicolumn{3}{c}{SNR $[-20,-10)$} &
\multicolumn{3}{c}{SNR $[-10,0]$} \\
\cmidrule(lr){4-6} \cmidrule(lr){7-9}
\cmidrule(lr){10-12} \cmidrule(lr){13-15}
& & &
SI-SDR & DMOS & STOI &
SI-SDR & DMOS & STOI &
SI-SDR & DMOS & STOI &
SI-SDR & DMOS & STOI \\
\cmidrule(lr){1-3} \cmidrule(lr){4-9} \cmidrule(lr){10-15}

\multirow{3}{*}{CRNN} & & -
& 1.54 & 1.98 & 0.36 
& 8.10 & 2.55 & 0.68
& -6.21 & 1.85 & 0.25 
& 5.29 & 2.39 & \textbf{0.55} \\
& 0.46 & Wiener
& 1.86 & 2.04 & 0.39
& 8.01 & 2.56 & 0.68
& -6.75 & 1.84 & 0.25
& 4.92 & 2.38 & 0.54 \\
& & cMPDR
& \textbf{5.55} & \textbf{2.31} & \textbf{0.46}
& \textbf{9.47} & \textbf{2.70} & \textbf{0.70}
& \textbf{-5.64} & \textbf{1.88} & \textbf{0.26}
& \textbf{5.48} & \textbf{2.46} & \textbf{0.55} \\

\cmidrule(lr){1-3} \cmidrule(lr){4-9} \cmidrule(lr){10-15}

\multirow{3}{*}{ULCNet} & & -
& 3.67 & 2.14 & 0.48
& 9.28 & 2.74 & 0.72
& \textbf{-4.00} & 1.90 & 0.31
& \textbf{6.44} & 2.52 & \textbf{0.60} \\
& 0.68 & Wiener
& 3.11 & 2.12 & 0.47
& 8.83 & 2.71 & 0.72
& -5.09 & 1.87 & 0.30
& 5.81 & 2.46 & 0.58 \\
& & cMPDR
& \textbf{6.41} & \textbf{2.42} & \textbf{0.54}
& \textbf{9.62} & \textbf{2.80} & \textbf{0.76}
& \textbf{-4.00} & \textbf{1.91} & \textbf{0.32}
& 6.29 & \textbf{2.54} & \textbf{0.60} \\

\bottomrule
\end{tabular}
\end{table*}

\subsubsection{Real-world dataset}\noindent
For the real-world dataset, clean speech and noise signals from the 2020 DNS Challenge dataset are used to generate 100 hours of noisy mixtures at the same sampling rate and SNR range as in the synthetic experiments. Models trained on these DNS mixtures are evaluated either on held-out DNS test data or on a second test set constructed from engine noise recordings of the IDMT-ISA Electric Engine dataset \cite{grollmisch_idmt-isa-electric-engine_2019}.
The DNS dataset mostly contains non-stationary noise types that are not specifically suited to the assumptions underlying cMPDR, allowing us to assess performance under less favorable conditions.
In contrast, the IDMT-ISA dataset exhibits pseudo-harmonic, cyclostationary noise. 
It is used as the primary real-world benchmark, as it is intentionally out-of-distribution relative to the training set and closely reflects the noise characteristics targeted by cMPDR.

\subsection{Baselines, performance metrics}\label{ssec:baselines-metrics}\noindent
The proposed method is compared with a ``no-preprocessor" baseline (the identity mapping $\Lambda_{\text{Id}}$) and with classic single-channel Wiener filter ($\Lambda_{\text{Wiener}}$) on the real-world datasets.
The Wiener filter assumes noise stationarity, with statistics estimated via the minimum statistics approach \cite{li_estimation_2015}.
It is worth noting that the DNN models were retrained for each preprocessing condition.
This ensures that each network is optimized specifically for its respective input distribution.

Performance was evaluated using four metrics.
Denoising performance is evaluated using scale-invariant signal-to-distortion-ratio (SI-SDR) \cite{roux_sdr_2019}.
Output quality is measured using the perceptual evaluation of speech quality (PESQ) \cite{rix_perceptual_2001} and overall DNSMOS \cite{reddy_dnsmos_2021}.
Intelligibility is assessed with the short-time objective intelligibility (STOI) metric \cite{taal_algorithm_2011}.

\subsection{Algorithm implementation and training details}\noindent
The CRNN was implemented in TensorFlow \cite{abadi2016tensorflow}.
It features three 2D convolutional layers with 8, 4, and 4 filters, and 3$\times$3 kernels. 
Each layer is followed by batch normalization and a max-pooling operation with a pooling factor of two, except for the first layer, which does not apply max pooling. 
The recurrent component is a GRU layer with 128 units, followed by a dropout layer with a dropout rate of 0.25. The final fully connected layer contains 256 units and Rectified Linear Unit (ReLU) activations are used throughout the network.
The ULCNet was reimplemented in TensorFlow in accordance with the specifications provided by the original authors \cite{shetu_ultra_2024}.
%

The datasets were partitioned into training, validation, and test sets using an 80/10/10 split.
STFT representations were computed with a $\SI{32}{\milli\second}$ analysis window, a $\SI{8}{\milli\second}$ hop size, and a 512-point FFT.
All models were trained using the Mean Absolute Error (MAE) computed in the time-frequency domain as the loss function and the Adam optimizer, with an initial learning rate of $1 \times 10^{-3}$. Gradient clipping was applied with a maximum norm of 3.0, and a learning rate scheduler reduced the learning rate by a factor of 0.5 if the validation loss did not improve for 3 consecutive epochs. Training was performed for a maximum of 300 epochs using 5-second long sequences.
Early stopping was employed to terminate training if the validation loss failed to improve for five consecutive epochs, and the model achieving the lowest validation loss was selected for evaluation on the test sets.

\section{Results}\noindent
\subsection{Evaluation on synthetic data}\noindent
\begin{figure}[tbph]
  \input{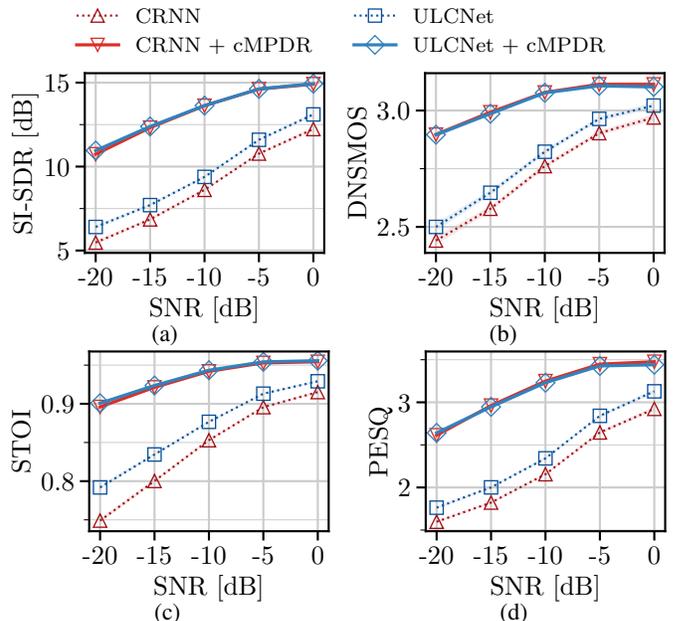}%
  \caption{Mean values of the objective metrics as a function of the SNR using the synthetic harmonic noise dataset. }
  \label{fig:simu}
\end{figure}
We first evaluate the pipeline on synthetic cyclostationary noise (\cref{ssec:synthetic_data}) to isolate the effect of cMPDR across different DNN architectures.
\cref{fig:simu} compares DNN-only baselines (dotted lines) with cMPDR-preprocessed pipelines (solid lines).
Three trends emerge from \cref{fig:simu}.
First, in the absence of preprocessing, ULCNet consistently outperforms the smaller CRNN.
Second, cMPDR provides substantial gains, particularly at low SNRs, where it increases CRNN performance by more than $\SI{6}{\decibel}$ SI-SDR and $0.15$ STOI, surpassing the standalone ULCNet by a large margin.
This indicates that incorporating domain-specific priors is more beneficial than increasing model capacity for this task.
Finally, the performance difference between cMPDR-preprocessed and DNN-only pipelines decreases at higher SNRs, indicating that cyclostationarity-aware preprocessing is most effective in strongly noise-dominated conditions.
\subsection{Evaluation on real-world data}\noindent
The second phase of the experimental evaluation focuses on assessing the impact of the cMPDR as a preprocessing stage across different DNN architectures on real-world recordings. 
These configurations are compared against two baselines: the same DNNs without preprocessing and with Wiener-filter preprocessing. \Cref{tab:enhancement_comparison} reports the objective performance results for the various models and preprocessing stages on each test set. For further analysis, the results are reported separately for a very low SNR range of [-20,-10) \si{\decibel} and a low SNR range of [-10,0] \si{\decibel}. Although the PESQ metric was computed, it is not reported in the table for readability. 
The PESQ scores exhibit the same trends as the other metrics, while overall speech quality is reflected by the DNSMOS metric.

First, we evaluated performance on the IDMT dataset, which consists of various engine noises that serve as examples of pseudo-harmonic and cyclostationary noise, attributes explicitly targeted by the cMPDR. \Cref{fig:spectrograms} presents spectrograms of a noisy sample from the IDMT test set, along with the corresponding clean target and the outputs of the CRNN model with and without cMPDR preprocessing. As illustrated in the figure, the cMPDR facilitates improved reconstruction of the target speech, particularly in frequency bands dominated by strong noise components.

Consistent with the observations from the synthetic noise experiments, the objective results reported in \Cref{tab:enhancement_comparison} for the IDMT test set indicate that cMPDR preprocessing achieves the highest scores across all evaluation metrics, irrespective of the evaluated DNN architecture, outperforming both the no-preprocessing and Wiener-filter baselines. Moreover, the advantages of the cMPDR become increasingly pronounced as the SNR decreases. In particular, under very low SNR conditions, the cMPDR yields improvements of up to \SI{3}{\decibel} in SI-SDR and 0.3 in DNSMOS relative to the best alternative preprocessing method for both models.
%
Notably, applying cMPDR to a simple and lightweight architecture such as the CRNN results in superior performance compared to using the more complex ULCNet without preprocessing or with Wiener filtering. This effect is reflected by gains of \SI{1.9}{\decibel} in SI-SDR and 0.15 in DNSMOS when comparing the cMPDR-CRNN configuration to the unprocessed ULCNet, under very low SNR conditions.

Second, we evaluated the same DNN architectures and preprocessing methods on the DNS test set. 
As shown on the right side of \Cref{tab:enhancement_comparison}, the cMPDR does not degrade performance for either architecture when processing mostly non-cyclostationary noise content, while providing modest but consistent improvements for the smaller CRNN model across all metrics.

Finally, it is worth noting that the results reported in \Cref{tab:enhancement_comparison} suggest that, under the considered SNR conditions, employing a Wiener filter as a preprocessing step can yield modest performance gains when the underlying architecture is sufficiently lightweight, as observed for the CRNN model. 
In contrast, for more complex architectures such as ULCNet, Wiener-filter preprocessing can be detrimental, with the no-preprocessing configuration achieving superior performance over Wiener-filter preprocessing.

\section{Conclusion}\noindent
This paper presented a hybrid speech enhancement framework combining a model-based cyclic minimum power distortionless response (cMPDR) preprocessor with supervised DNN-based denoising.
By explicitly leveraging cyclostationarity, the proposed approach improves speech enhancement performance in low-SNR conditions dominated by rotating machinery noise.
Experimental results show that cMPDR preprocessing allows low-capacity DNNs to outperform models with 30\% more parameters operating on raw or Wiener-filtered inputs, highlighting the benefit of incorporating signal structure that is difficult for DNNs to learn implicitly.
The cMPDR is distortionless, so in the absence of cyclostationary interference, the preprocessing stage does not degrade the target speech. 
Its effectiveness relies on noise frequencies being stable and accurately estimated.
Future work could enable DNNs to directly leverage cyclostationarity, for example by conditioning on frequency information or using frequency-shifted representations.

\bibliographystyle{IEEEtran}
\bibliography{IEEEabrv,references}

\end{document}